# Bioavailability of Ag(I) with *Arthrobacter oxidas and Arthrobacter globiformis*

*E.Gelagutashvili, A.Rcheulishvili, E.Ginturi, N.Bagdavadze, N.Kuchava , K.Tsakadze, M.Janjalia*

*I. Javakhishvili Tbilisi State University*
*E. Andronikashvili Institute of Physics*
*0177, 6, Tamarashvili St.,*
*Tbilisi, Georgia*

## Abstract

The biosorption of Ag(I)_ *Arthrobacter* species (*Arthrobacter globiformis 151B* and *Arthrobacter oxidas 61B*) was studied at simultaneous application of dialysis and atomic absorption analysis.

The biosorption constants and nature of interaction for Ag(I) –*Arthrobacter oxidas* and Ag(I) –A*rthrobacter globiformis* were determined.

The biosorption constants for Ag(I)- –A*rthrobacter globiformis* and for Ag(I) – *Arthrobacter oxidas* equal to 65.0 $x10^{-4}$ and 35.0 $x10^{-4}$ respectively.

## Introduction

Microorganisms have evolved various mechanisms to resist the heavy metal induced stress. These mechanisms include accumulation and complexation of the metal ions inside the cell, the biosorption etc.

The search for new technologies, involving the removal of toxic metals from wastewaters has directed attention to biosorption due to the metal-binding capacities by various biological materials [1]. Metal sorption performance depends on some external factors such as pH, organic materials in solution, temperature etc. Paper [2] deals with silver sorption by *Myxococcus xanthus* biomass. The dry biomass of these microorganisms has shown to be a good sorbent for the recovery of silver present at low solution concentrations. Study silver uptake by *Pseudomonas dimunuta* was carried out by growing the bacteria in a chloride-free medium containing silver ions [3]. *Arthrobacter oxydans* P52 isolated from soil samples was found to degrade the phenylcarbamate herbicides phenmedipham and desmedipham cometabolically by hydrolyzing their central carbamate linkages[4]. Cell-free culture supernatants of five psychrophilic bacteria, *Arthrobacter gangotriensis* and others, have been used to synthesize silver nanoparticles (AgNPs )[5]. There is a sharp increase of the interest of studying the interaction of silver ions with microorganisms.

The objective of this study was to investigate the biosorption of silver ions by *Arthrobacter species (Arthrobacter oxidas 61B and Arthrobacter globiformis 151B)* using methods of dialysis and atomic absorption analysis.

## Materials and Methods

*Arthrobacter* bacteria were cultivated in the nutrient medium [6]. Cells were centrifuged at 12000 rpm for 10 min and washed three times with phosphate buffer (pH 7). The centrifuged cells were dried without the supernatant solution until constant weight. After solidification (dehydrated) of cells (dry weight) solutions for dialysis were prepared by dissolving in phosphate buffer. This buffer was used in all experiments. $AgNO_3$ – Analytical grade.

The study of biosorption of Ag(I)-*Arthrobacter oxidas* 61B *and* Ag(I)- *Arthrobacter globiformis* 151B was carried out by the methods of dialysis and atomic absorption analysis. *Arthrobacter oxidas* and *Arthrobacter globiformis* were dissolved in phosphate buffer, pH 7.0. The dialysis experiments were carried out in 5ml cylindrical vessels made of organic glass. A cellophane membrane of 30μm width (type - "Visking" manufacturer - "serva") was used as a partition. The duration of dialysis was 72 hours. The experiments were carried out at $23^0$C temperature. In all mentioned cases, the concentrations of *Arthrobacter oxidas* and *Arthrobacter globiformis* were 1 mg/ml. The metal concentration ranged from $10^{-2}$ to $10^{-5}$. The metal concentration after the dialysis was measured by atom-emission analysis at $\lambda=328.1$ *nm* wavelength. Each value was determined as an average of three independent estimated values ± the standard deviation. The processing of experimental data was performed using different models. Using of number of mathematical models has revealed that the Hill model [7] is the most accurate approximation.

Results and discussions

In Fig. 1, there are presented the biosorption isotherms, when the *Arthrobacter oxidas and Arthrobacter globiformis* are dissolved in phosphate buffer *pH 7.0*. The dots on the figure show the experimental data, and the solid line in both cases is the hypothetical theoretical curve chosen by $\chi^2$ criterion ($\chi^2$ *0.2*, $R^2$ *0.97*) in $Y=C_b/n$ *vs Logm* coordinates, where the number (*n*) of active centers (capacity) is determined as the maximum value of $C_b$; $C_b$ is concentration of metal ions adsorbed and *m* is the concentration of free metal ions. Each dot is the average of three independent values, and the standard deviation < 14% of average value. It is seen from fig.1 that the type of *y* vs *logm* dependence is nonlinear – *S*-shape. This means that there exists positive cooperation of interaction of metal ions bound with *Arthrobacter oxidas* and *Arthrobacter globiformis* , i.e. binding of the first metal ion increases affinity of the site for the second one. For more argumentation, as an additional criterion of cooperativity, the Hill model was used [7]. The linear form of the Hill equation is equal to:

$$\log Y/1-Y=\log K+1/n_H \log m,$$

where y and m are the same, as in fig.1; *K (biosorption constant)* and $n_H$*(Hill coefficient)* are empirical constants and have been calculated from intercept and slope of the plots. Taking this into account, the same data are presented in Hill coordinates, (Fig. 2).

By using Hill equality, the values of *K* and $n_H$ were obtained. These values in addition to R and standart deviation values are shown in table 1. In both cases, the correlation between the experimental and the theoretical data is obvious (*R* is more than 0.9). The result obtained from the cooperative interaction shows that the biosorption constants for Ag(I) - *Arthrobacter oxidas and* Ag(I)- *Arthrobacter globiformis* complex and the capacity depend on the species of bacteria. In particular, the biosorption characteristics are $K=65x10^{-4}$ and n=2.1 for *Arthrobacter globiformis* and $K=35x10^{-4}$ and n= 2.7 for Ag(I) - *Arthrobacter oxidas*.

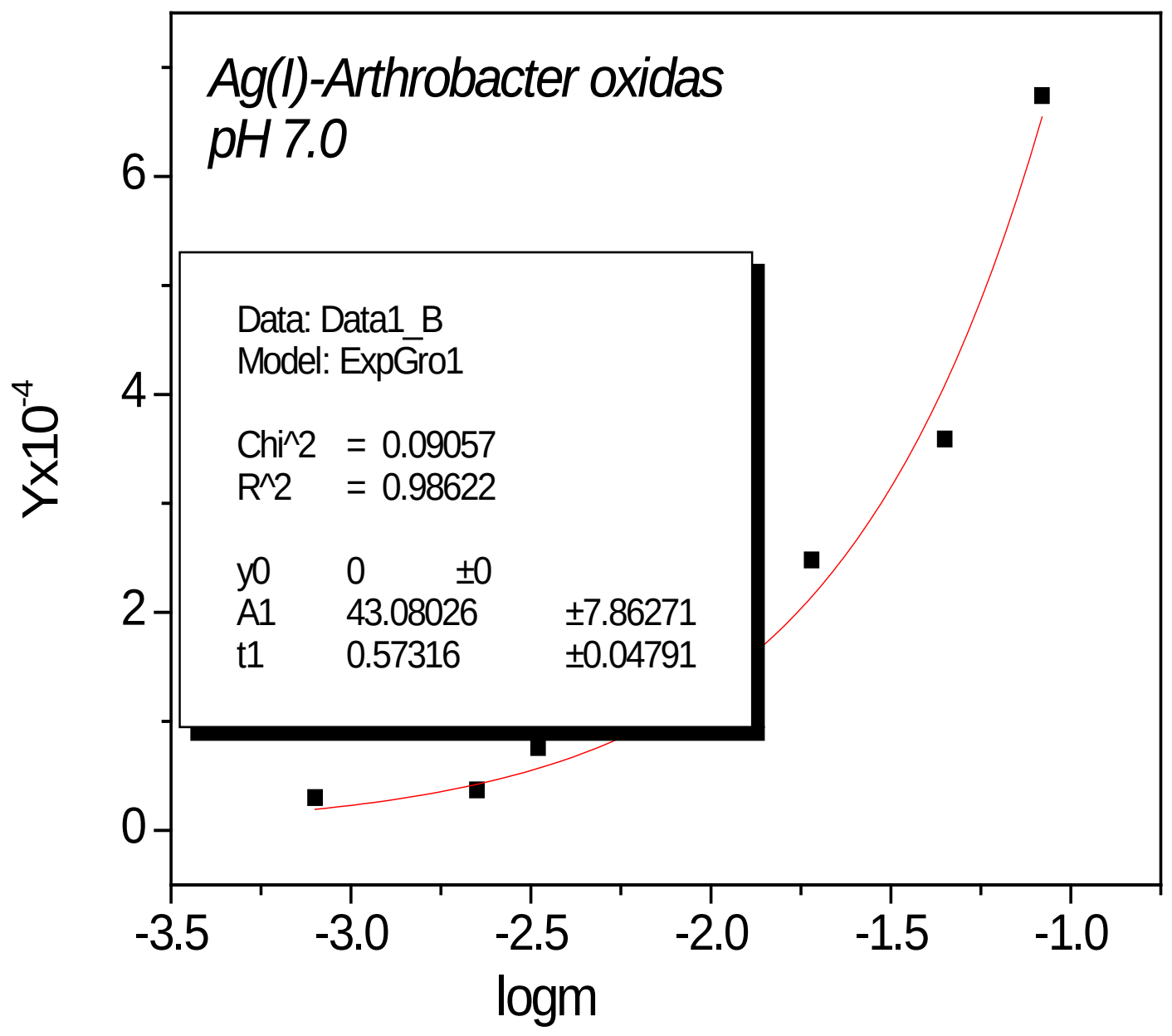

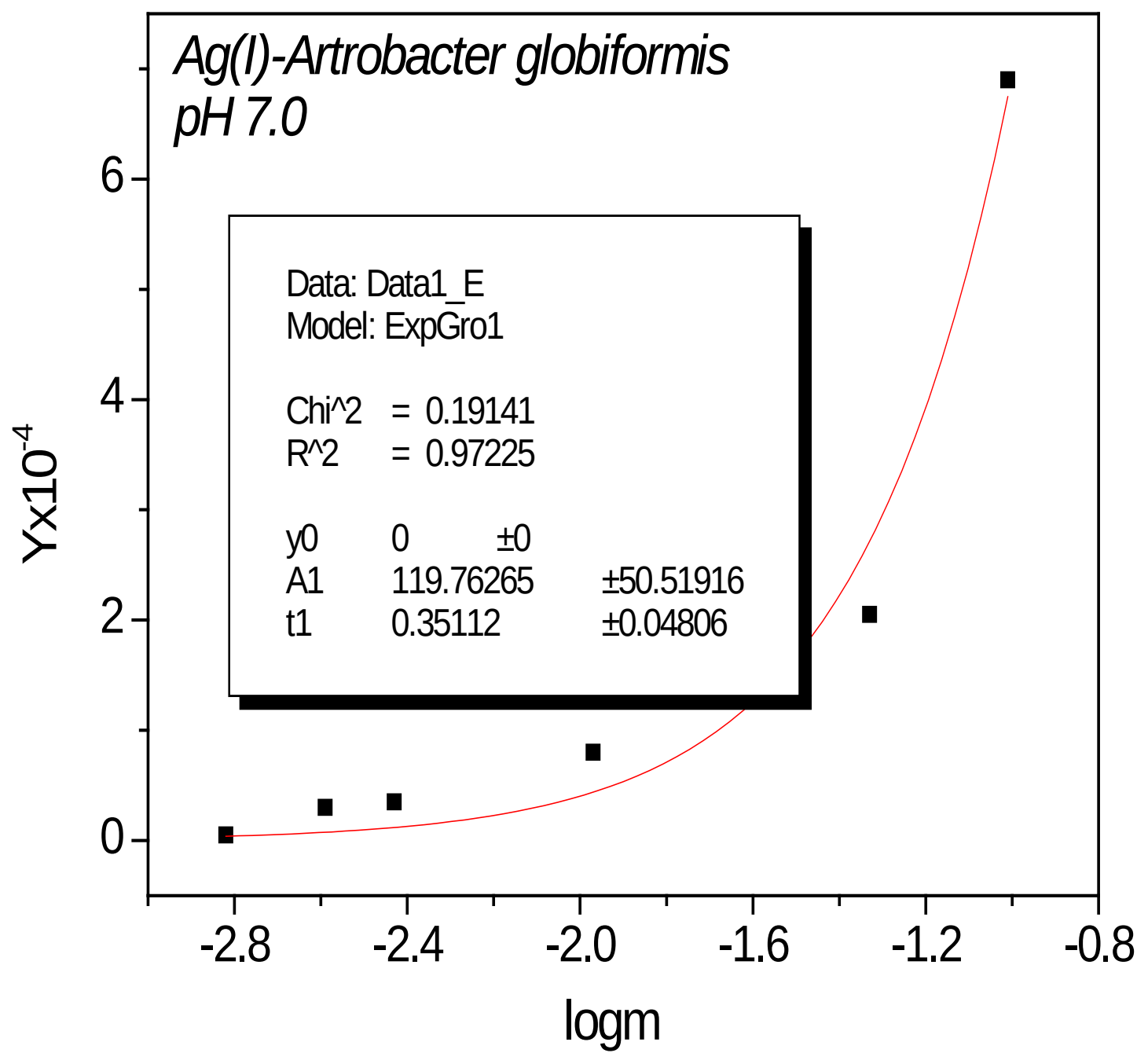

Fig. 1. Biosorption isotherms for Ag(I) - *Arthrobacter oxidas* *and* Ag(I)-*Arthrobacter globiformis* complex for neutral *pH* (dissolved in phosphate buffer).(in figures are shown the hypothetical theoretical curves chosen by $\chi^2$ criterion ($\chi^2$ 0.2).

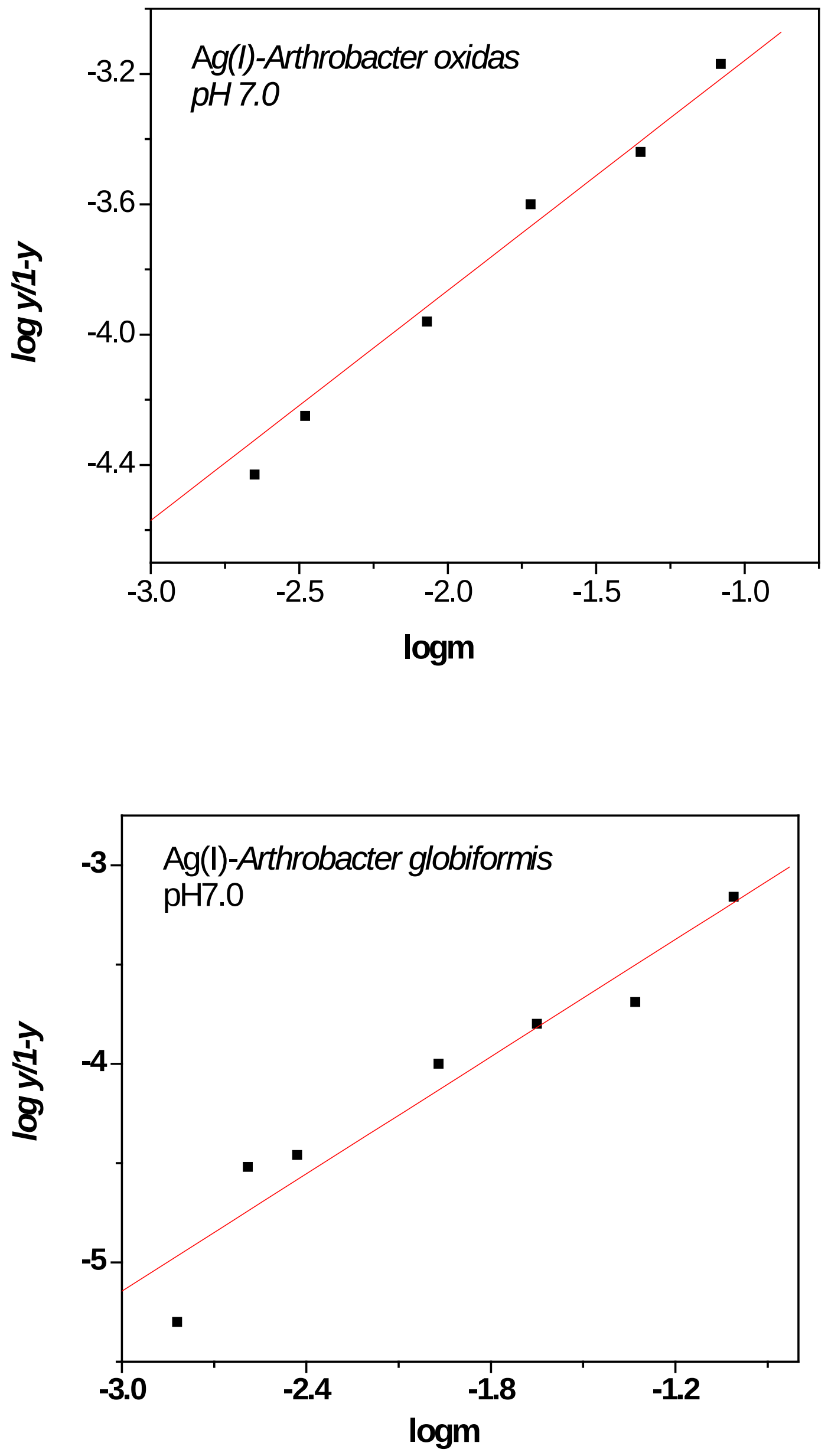

Fig. 2. Biosorption isotherms for Ag(I) - *Arthrobacter oxidas* and Ag(I)- *Arthrobacter globiformis* complex for neutral *pH* in Hill coordinates.

Table 1. Biosorption parameters for *Ag(I) - Arthrobacter oxidas and Ag(I)- Arthrobacter globiformis* complex for neutral *pH*

| | | *Arthrobacter oxidas 61B* | *Arthrobacter globiformis 151B* |
|---|---|---|---|
| Biosorption constant | *K* | 35.0 $x10^{-4}$ | 65.0 $x10^{-4}$ |
| Biosorption capacity | *n* | 2.7 | 2.1 |
| Hill coefficient | $n_H$ | 1.4 | 1.02 |
| Standard deviation | *SD* | 0.09 | 0.21 |
| Correlation coefficient | *R* | 0.98 | 0.96 |

Hence, the biosorption constant of Ag(I)- *Arthrobacter globiformis* is more, than that for Ag(I) - *Arthrobacter oxidas*. The capacities are also different. It is clear, that biosorption constants in both cases are quite high. Gram positive bacteria cell wall consists of a variety of polysaccharides and proteins and thus offers a number of active sites capable of binding metal ions. Consequently, gram-positive bacterial cell surface is characterized by stronger bonding with positively charged silver ions. Gram-positive bacteria have a greater sorptive capacity due to their thicker layer of peptidoglycan which contains numerous sorptive sites [8]. It was shown also, that the carboxyl groups were the main binding sites in the cell wall of gram positive bacteria [9]. It was found in [3] that higher amounts of silver ions were accumulated inside the cell during early exponential phase compared to the amount bound on the cell surface from *P. dimunuta.* In [2], it was shown that first: silver was bound to the cell surface and extracellular polysaccharide, and second: a silver intracellular deposition process took place. Transmission electron microscopy study of *M. xanthus* wet biomass after silver accumulation shows the sorption within the extracellular polysaccharide on the cell wall, and in the cytoplasm. On the other hand, the binding characteristics of metal cations during biosorption can partially be explained by Pearson' s concept of hard and soft acid and base theory and by Irving-Williams series. Silver is soft metal and soft metals involve mainly covalent or coordinative interaction with sulfur or nitrogen –containing ligands[10].

As is seen from Table 1, $n_{H/}n<1$, showing the incomplete cooperativity of interaction between the silver ions and *Arthrobacter oxidas and Arthrobacter globiformis*. The mentioned model gives the possibility to characterize the nature of interaction of Ag(I)- *Arthrobacter oxidas* and - *Arthrobacter globiformis*.

Thus, using Hill models, it is possible to determine not only biosorption constant and capacity for Ag(I) - *Arthrobacter oxidas and* Ag(I)- *Arthrobacter globiformis* complex, but to characterize the nature of interaction as well.

This work was supported by grant STCU #4744

## R E F E R E N C E S